\documentclass[
amsfonts,amssymb,aps,prb,groupedaddress,showpacks,
a4paper,floatfix]{revtex4-1}

\usepackage{graphicx}
\usepackage{epsfig}
\usepackage{bm}
\usepackage{natbib}
\usepackage{revsymb4-1}

\begin{document}


\topmargin=1pt

\title{Temperature evolution of superparamagnetic clusters in single-crystal La$_{0.85}$Sr$_{0.15}$CoO$_3$ from nonlinear magnetic ac response and neutron depolarization}

\author{A.~V.~Lazuta}
\author{V.~A.~Ryzhov}
\author{V.~V.~Runov}
\author{V.~P.~Khavronin}
\author{V.~V.~Deriglazov}
\email[]{deriglaz@pnpi.spb.ru}
\affiliation{NRC \textquotedblleft Kurchatov Institute\textquotedblright~PNPI, Gatchina, Russia}


\begin{abstract}
The representative measurements of the second harmonic in ac magnetization complemented by neutron depolarization have been performed for single-crystal La$_{0.85}$Sr$_{0.15}$CoO$_3$ in the temperature range 97 K$<T<230$ K, where occurrence of a small fraction ($\alt 10^{-3}$) of nanoscale ferromagnetic clusters (FMC) has been found. Magnetic, geometrical and dynamical parameters of the FMC system have been evaluated in the temperature range $T<140$ K, where superparamagnetic regime installs, by means of the formalism involving the Fokker-Planck equation (FPE). With lowering the temperature, the amount of clusters fraction, the cluster size and magnetic moment along with its diffusion relaxation time strongly increase, each in its own temperature interval. Below 130 K, FMC contribute essentially to the total linear magnetic susceptibility. The damping factor of the order $10^{-1}$ proves the importance of precession in thermal relaxation of the cluster magnetic moment. The FMC are a precursor of long-range ferromagnetic correlations seen below 100 K with neutron-scattering techniques. The employed technique supplemented with FPE-based data-treatment formalism is a novel method for studying superparamagnetic systems.

\end{abstract}



\maketitle

\section{Introduction}

Hole-doped lanthanum cobaltites are known to exhibit the tendency to magnetoelectronic phase separation (MEPS) in the wide range of doping and temperature [\onlinecite{r6,r7,r11,r12,r33,r35}]. The parent compound LaCoO$_3$ in the ground state is the diamagnetic insulator. The $3d$ orbital of Co$^{3+}$ ions is split by nearly cubic crystal field, slightly exceeding intra-ion exchange, in such a way that $t_{2g}$ orbital triplet is completely filled with six electrons, forming the low-spin (LS) state with the spin $S=0$, whereas the empty $e_g$ orbital doublet lies 10-12 meV higher. At elevated temperatures, LaCoO$_3$ becomes paramagnetic due to the thermally induced transition to higher spin states.

The hole doping with alkali-earth ions Sr$^{2+}$ gives rise to nanoscale MEPS in the form of spin polarons and/or larger spin clusters depending on temperature and doping, as established from neutron diffraction, nuclear magnetic resonance (NMR), M{\"o}ssbauer spectroscopy [\onlinecite{r37}], small-angle neutron scattering, inelastic neutron scattering, x-ray spectroscopies, heat capacity, magnetometry, ac magnetic susceptibility (MS), thermopower, transport and magnetotransport measurements (Refs.~[\onlinecite{r32,r33}] and a wide bibliography therein). On the \textit{x-T} phase diagrams [\onlinecite{r6,r38}], the MEPS state covers a broad area below the paramagnetic phase. The transition temperature increases with doping, so that, e.g. at $x=0.15$, the spin-polaronic state extends up to $T\approx 250$ K [\onlinecite{r6}]. At lower temperatures somewhat below 60 K, a highly inhomogeneous state develops [\onlinecite{r6,r7}], its particular form depending on $x$. For the doping less than the characteristic value $x_c=0.17$, pronounced hole-rich and hole-poor regions coexist in the form of insulating spin glass, whereas in the doping range $x_c<x<0.25$, the spin-polaronic state evolves upon cooling followed by the long-range-order ferromagnetic (FM) region eventually passing to the metallic phase of strongly inhomogeneous ferromagnet [\onlinecite{r6}]. According to other sources (Ref.~[\onlinecite{r32}] and references therein), the insulator-metal transition occurs in the interval 0.18-0.22, with the percolation onset at $x_c=0.18$.

The nature and temperature evolution of MEPS close to $x_c$ has been of keen interest. In the insulator-state vicinity of $x_c$, large-scale magnetic inhomogeneities in single crystals have been known to occur at temperatures not exceeding 100 K implying spin-polaronic MEPS at higher temperatures [\onlinecite{r6}]. An extensive neutron scattering study of the single-crystal La$_{0.85}$Sr$_{0.15}$CoO$_3$ revealed coexistence of the regions with long-range and short-range order at temperatures $T<100$ K [\onlinecite{r7}]. The former were identified as FM clusters of the diameter larger than 7 nm and the latter as referring to smaller, less than 2 nm, entities. Under external magnetic field, large clusters formed at the expense of small ones. After switching off the field, some part of the large clusters still persisted. The small-angle neutron scattering assisted by magnetotransport measurements for the same composition also showed spontaneous MEPS developing below 150 K in the form of small FM clusters embedded in a non-FM matrix reaching the size 1.5-2.5 nm at low temperatures [\onlinecite{r12}].

It is generally accepted that, similarly to manganites, in doped cobaltites double exchange interaction between Co$^{3+}$ and Co$^{4+}$ leads to formation of FM clusters with metallic conductivity with higher density of charge carriers than in the surrounding matrix. Increase of the Coulomb energy with the cluster growth is compensated by kinetic energy of the carriers up to some critical cluster size, above which the clusters become unstable [\onlinecite{r47,r48,r49,r50}].

The MEPS occurs in manganites, as well, revealing a certain resemblance to that in cobaltites at a close level of doping, in compounds both conserving their insulator state and exhibiting insulator-metal transition below the Curie temperature [\onlinecite{r10,r51}]. In manganites, however, the increase of paramagnetic fluctuations upon cooling resulting in the FM phase transition impedes and restricts in temperature observation of FM clusters [\onlinecite{r10}]. The latter is easier for cobaltites as their matrix is only slightly paramagnetic down to lower temperatures.

In polycrystalline LaSr cobaltites, FM clusters occur in the wider doping and temperature ranges, as crystallite boundaries favor their formation. From linear and non-linear magnetic susceptibility measurements [\onlinecite{r23}], FM clusters were found to exist at the dopings $x>0.25$ and to appreciably contribute well above 100 K even in the insulating state for $x=0.1$ and 0.15. 

Previously, the non-linear magnetic response was found to occur in single-crystal La$_{0.85}$Sr$_{0.15}$CoO$_3$ in the wide temperature range below $T\approx 213$ K [\onlinecite{r13}]. The preliminary qualitative analysis has shown that, below this temperature, a small fraction of FM clusters emerges, exhibiting nonsuperparamagnetic behavior characterized by considerable field hysteresis. Localization of the clusters was attributed to preferable sites formed by slight doping heterogeneity and local oxygen nonstoichiometry. The concentration of clusters increased weakly with lowering the temperature. On cooling below $T\approx 147$ K, homogeneous nucleation of FM clusters developed exhibiting superparamagnetic (SPM) behavior with only a small field hysteresis. Starting from 135 K, the concentration of SPM clusters increased intensively in a narrow temperature interval.

However, the former studies [\onlinecite{r8,r9,r10}] provide quite limited, indirect, and approximate information on FM clusters being based on the analysis of the response signal itself, viz., extrema positions, signal amplitudes, field asymptotes, etc. In this report, we specify the previous observations and present a detailed complete quantitative characterization of the cluster system involving a thorough mathematical formalism for data treatment especially focusing on the temperature region below 140 K. Evolution of clusters was traced down to 97 K at magnetic fields not exceeding 300 Oe. From representative data on the second harmonic of longitudinal magnetic response complemented by neutron depolarization results, a large set of parameters characterizing geometrical, magnetic, and dynamical features of the system of FM clusters in SPM regime was determined using the formalism based on the solution of the Fokker-Planck equation (FPE) [\onlinecite{r1,r2,r3,r4,r5}]. With these parameters, we visualize the above mentioned successive phases of the cluster MEPS, specify the temperature boundaries of the MEPS stages, and outline two different regimes of homogeneous nucleation alternating each other at 115 K.

The observed large SPM clusters are suggested to be (i) a precursor of the FM state occurring below the insulator-metal transition at the doping $x>x_c$ and (ii) an onset of large FM clusters revealed by Phelan and collaborators with neutron scattering techniques at lower temperatures, $T<100$ K [\onlinecite{r7}].

The technique employed in this research, though not being conventional, proved its high efficiency in studying large, of the order 10 nm, FM particles [\onlinecite{r8,r9,r10}]. Such objects appear to give the main contribution to generation of the second harmonic at magnetic fields of the order 10-100 Oe due to strong nonlinearity of their response in weak fields. In this study, the signal from the SPM fraction occupying the sample volume $\agt 10^{-4}$ was reliably measured and the parameters characterizing the SPM system were obtained with sufficient accuracy. The ac-field frequency matched the most informative frequency range where the real and imaginary parts of the second-order response signal were comparable.

In Sec. II, relevant experimental details both for ac and neutron depolarization measurements are presented. Section III acquaints with the principles of extracting the quantities characterizing the cluster system and explains how some more parameters can be evaluated involving additionally the neutron depolarization data. In Sec. IV, the neutron depolarization data and the non-linear response for some typical temperatures are presented and shortly described. Temperature dependencies of the characteristic parameters are analyzed and a scenario of the cluster-system evolution is discussed with the separate emphasis on the SPM dynamics. The linear ac susceptibility retrieved from the non-linear data is compared to the measured linear response. Section V is devoted to quite a special problem concerning the grounds of SPM dynamics. Experimental data on the non-linear ac response appear to be representative enough to distinguish between Landau-Lifshitz and Gilbert ansatz underlying the Fokker-Planck formalism and to make a choice in favor of the latter. In the conclusion section, the main results are summarized. The appendix contains the formalism for ac nonlinear data treatment.

\section{Experimental details}

The manufacturing and certifying of the specimen are described in detail in Ref.~[\onlinecite{r13}]. The feed rod for the sample was fabricated with the standard routine based on solid-state synthesis. The single-crystal La$_{0.85}$Sr$_{0.15}$CoO$_3$ was grown by the floating-zone technique using radiative heating under an oxygen pressure of 1 bar with a typical growth rate of 1 mm/h. The crystal was found to be single-phase, pseudocubic with the slight rhombohedral distortion at room temperature (the space group $R{\bar 3}c$), as confirmed by x-ray diffraction. The element content was monitored by an x-ray microanalyzer with the relative accuracy 3\%.

The second harmonic of magnetization $M_2$ was measured in \textit{parallel} dc and ac magnetic fields on the setup [\onlinecite{r29}] approved in a good deal of magnetic measurements (Refs.~[\onlinecite{r8,r9,r10,r14,r15}] and references therein). The dc field $H$ was scanned symmetrically with respect to zero field within $\pm 300$ Oe with the round-up cycle 0.14 s. A frequency and an amplitude of the excitation field were $\omega/2\pi=15.65$ MHz and $h=0.8-14$ Oe, respectively. The latter was optimized to obey the condition $M_2\propto h^2$, necessary for the preliminary comparative data analysis, although not mandatory in the FPE data-treatment formalism. Both phase components of the signal, Re$M_2$ and Im$M_2$, were simultaneously recorded as functions of the dc field in the temperature region 97 K $<T<230$ K. The sample temperature was stabilized with an accuracy of 0.2 K.

The neutron depolarization technique is known to be a powerful
mesoscopic technique to study ferromagnetic clusters in a phase
separated system [\onlinecite{r43,r44,r45,r46}]. The neutron magnetic moment interacts with the internal magnetic field of an FM cluster. As the internal fields of different clusters are thermally disordered, the neutron beam, passed through the cluster ensemble, depolarizes. The contribution of spin polarons to neutron depolarization via forward scattering is negligibly small, due to their small size. The scattering on structural inhomogeneities also does not affect the neutron depolarization. However, in the case of too small concentration of clusters and/or the cluster size, the depolarization can be also small, demanding high precision of the depolarization measurements.

The neutron depolarization was measured at the small-angle polarized-neutron facility \textquotedblleft Vector\textquotedblright (the reactor WWR-M, Gatchina, Russia) [\onlinecite{r30}] in the temperature range 50-290 K, covering the relevant temperatures. The polarized neutron beam with the wavelength $\lambda=0.85$ nm and the spectrum half-width $\Delta\lambda/\lambda=0.25$ passed through the cylinder sample 3.8 mm in diameter and 13.8 mm high, that was cut from the same parent single crystal, from which the sample for the ac measurements was cut off. A diaphragm at the sample was 2$\times$12 mm$^2$. The guiding magnetic field 5 Oe at the sample had a minor effect on the clusters state, if at all. To ensure the absence of temperature hysteresis in the $T$-range under study, the measurements were performed both with decreasing and increasing the temperature. The measurement time 5 h at each temperature point was sufficient for the accuracy needed.

The linear ac MS was measured, as well, in zero dc field with the ac field of the frequency 95 KHz and the amplitude 1 Oe in the wide temperature range 80 K $<T<180$ K, above the freezing temperature.

\section{Details of data treatment}

The experimental data treatment was performed at the PNPI computer cluster.

The $M_2$ measurement data were fitted with the model function obtained from the stationary solution of FPE [\onlinecite{r3,r4}],
\begin{equation}
2\tau_N\frac{\partial W}{\partial t}=\frac{\beta}{\alpha}{\bf u}\cdot(\nabla V\times\nabla W)+\nabla(\nabla V+\beta W\nabla V)~,\label{one}
\end{equation}
derived from the underlying Landau-Lifshitz-Gilbert stochastic equation [\onlinecite{r17a}] for the magnetization vector of a single-domain particle. In Eq.~(\ref{one}), $W$ is the nonequilibrium probability-density function for directions ${\bf u}={\bf M_s}/M_s$ of the magnetization ${\bf M_s}$, $\nabla=\partial/\partial{\bf u}$ is the gradient operator, $t$ is time, $V$ is the magnetic potential, and $\beta=v/k_BT$, where $v$ is the particle volume, and $k_B$ and $T$ are the Boltzmann constant and the temperature, respectively. The diffusion relaxation time was taken in Landau-Lifshitz form as $\tau_N=\tau_\circ/\alpha$, where $\alpha$ is the damping factor and $\tau_\circ=\beta M_s/2\gamma$ with $\gamma$ being the gyromagnetic ratio.

The magnetic potential $V$ is assumed to be \textit{uniaxial} [\onlinecite{r1,r2,r3,r4}], viz.,
\begin{equation}
\beta V=\sigma\sin^2\vartheta-\xi_H{\bf u}\cdot\frac{{\bf H}}{H}-\xi_h{\bf u}\cdot\frac{{\bf h}}{h}\cos\omega t~.\label{two}
\end{equation}
The first term in Eq.~(\ref{two}) is the anisotropy energy with $\sigma=\beta K_a$ where $K_a$ is the anisotropy constant and $\vartheta$ is the angle between the magnetization vector and the anisotropy axis. The second and the third terms are the Zeeman energies of the particle magnetic moment in the bias dc magnetic field ${\bf H}$ and the ac driving magnetic field ${\bf h}$ of the frequency $\omega$, respectively, with $\xi_H=\beta M_sH$, $\xi_h=\beta M_sh$. No restriction on the sign of $\sigma$ is implied.

In the framework of the formalism, the time scaling factor $\tau_N$ in Eq.~(\ref{one}) denotes the characteristic time of diffusion in the absence of potential (free-diffusion time). In the case $\sigma>0$, it is a pre-exponential factor in the conventional N{\'e}el relaxation time derived for a double-well potential with a barrier.

The first and the second terms in the right-hand side of Eq.~(\ref{one}) are responsible for precession and thermal relaxation, respectively. When the magnetic field is parallel to the anisotropy axis, the problem simplifies considerably [\onlinecite{r17a}]. In the present study, however, this was not the case. First, by technological reasons, the anisotropy axis was oriented by the angle $\vartheta\approx 41.4^\circ$ relative to the sample plane and, hence, to the applied in-plane magnetic fields. Second, the point Laue patterns indicated some twinning in the single crystal used, usually observed in pseudocubic perovskites [\onlinecite{r19}], necessitating averaging over the twins orientations. And third, the axial symmetry, if conserved, degenerates precession, eliminating the precession term in Eq.~(\ref{one}), what strongly diminishes the informative content of the data. An analytical solution of Eq.~(\ref{one}) for the common case of oblique magnetic field is absent, and one needs to solve the problem numerically. However, by expanding the solution in series by spherical harmonics, it is still possible to reduce the problem to an infinite system of equations, which, in turn, can be expressed as a continued-fraction matrix relation (see appendix).

An experimental geometry with the parallel orientation of dc- and linearly polarized ac magnetic fields somewhat simplifies the data treatment, with no damage for completeness of the information content.

The solution accuracy depends, mainly, on the number of retained equations in the system (the number of iterates in solving the continued-fraction equation) and on the number of retained harmonics in the Fourier expansion. The former number ${\rm n}_\circ=8$ and the latter ${\rm k}_\circ=\pm 4$ appeared to be sufficient for the whole data array.

According to common symmetry requirements, the second harmonic must be antisymmetric on the dc magnetic field. However, due to the finite cycling frequency of $H$, hysteretic behavior may occur, violating the $H$-antisymmetry for a single, direct or reverse, $H$-scan, while the antisymmetry still conserves for the whole hysteresis loop. As the present stationary solution, with the steady field $H$, does not account for the hysteresis, an average between the direct and reverse scans was taken for each temperature, which, as expected, turned out to be antisymmetric within the experimental error. This average was additionally antisymmetrized relative to $H=0$. Such elimination of a small hysteresis was believed not to distort the resultant parameters considerably.

The real and imaginary parts were simultaneously fitted with the following parameters: the saturation magnetization of the cluster ensemble $M$, the anisotropy energy of a cluster $\varepsilon_a=v_cK_a$, where $v_c$ is the mean cluster volume, the mean magnetic moment of clusters $m_c=v_cM_s$, the dispersion of the log-normal distribution $D_v$ over cluster volumes, the damping factor $\alpha$ and two parameters describing the real and imaginary parts of the signal from the matrix assumed to be linear functions of the dc magnetic field. The matrix response, when observed, includes the signals coming from magnetic inhomogeneities other than SPM clusters, e. g., from spin polarons. Due to their small size, they exhibit only slight non-linearity in weak fields and make up a minor contribution to the total response in the measured region $H<300$ Oe, despite that these entities may occupy a large volume of the sample.

Some additional magnetic characteristics of the cluster system can be calculated from the fit parameters, as well. These are the concentration of clusters $N=M/m_c$, the mean intercluster distance $\langle r\rangle\propto N^{-1/3}$, the incluster anisotropy field $H_a=\varepsilon_a/m_c$, the diffusion relaxation time of the cluster magnetic moment $\tau_N=m_c/2\gamma\alpha k_BT$, and the characteristic dipolar energy associated with the cluster subsystem $\varepsilon_d=4\pi m_c^2/{\cal V}$, where ${\cal V}\propto N^{-1}$ is the mean volume per cluster evaluated from the concentration of clusters with the known perovskite-cube volume $v_\circ$.

With the neutron depolarization data additionally involved, some more quantities characterizing the SPM system can be obtained. The polarization of the passed neutron beam can be expressed, in notations of Ref.~[\onlinecite{r18}], as
\begin{equation}
P=P_\circ\exp\biggl\{-\frac{4}{3}\biggl(\frac{\gamma_nB}{V_n}\biggr)^2\mathcal{R}C^{1/3}L\biggr\}~, \label{three}
\end{equation}
where $P_\circ$ is the initial polarization directed along the beam, $\gamma_n$ is the neutron gyromagnetic ratio, $V_n$ is the neutron speed, so that $\gamma_n/V_n\approx 46.3\lambda$ nm$\cdot$Oe$^{-1}$, $B=4\pi\langle\mu\rangle/v_\circ$ is the magnetic induction in FM clusters with $\langle\mu\rangle$ being the mean magnetic moment per formula unit, $\mathcal{R}$ is the mean radius of clusters, $C$ is the weight (or volume) fraction of clusters in the sample and $L$ is the sample thickness along the neutron beam. The value $L=3.58$ mm was used as an average over the diaphragm width 2 mm.

Equation~(\ref{three}) makes it possible to separate out the incluster magnetic moment $\langle\mu\rangle$, the mean cluster volume $v_c$, and the fraction of clusters $C$. By definition, the mean cluster magnetic moment reads:
\begin{equation}
m_c=\frac{v_c}{v_\circ}\langle\mu\rangle~,\label{sysone}
\end{equation}
noting that $v_c/v_\circ$ is the mean number of formula units per cluster. The saturation magnetization $M$, a normalization factor for the $M_2$ response function in the units emu$\cdot$g$^{-1}$, can be presented as
\begin{equation}
M=C\langle\mu\rangle\frac{N_A}{{\mathcal M}_{mol}}~,\label{systwo}
\end{equation}
where $N_A$ and ${\mathcal M}_{mol}$ are the Avogadro number and the molecular weight, respectively. From Eq.~(\ref{three}), one obtains:
\begin{equation}
Cv_c\langle\mu\rangle^6=\frac{9\pi}{16L^3}\biggl(\frac{v_\circ V_n}{4\pi\gamma_n}\biggr)^6\ln^3\biggl(\frac{P_\circ}{P}\biggr)\exp\biggl(-\frac{3D_v}{2}\biggr)~.\label{systhree}
\end{equation}
Remind that $D_v$ is a dispersion of the volume distribution of the cluster ensemble. The exponential in Eq.~(\ref{systhree}) is a correction factor to transfer from the cluster volume with the radius ${\mathcal R}$ (Eq.~(\ref{three})), $v_{\mathcal R}=4\pi{\mathcal R}^3/3$, to the mean cluster volume $v_c$ to be obtained. This radius can be expressed as an average over the volume distribution in the form ${\mathcal R}\propto\int{\rm d}v\sqrt[3]{v} f(v)\tilde v/v$ where $\tilde v$ is a median of the assumed log-normal distribution $f(v)$. The term $\tilde v/v$ occurs since $f(v){\rm d}v$ is taken as a fraction of the \textit{total volume} occupied by the clusters with volumes in the interval $(v,v+{\rm d}v)$. With the same note, $v_c=\int{\rm d}vvf(v)\tilde v/v$ just equals to the median $\tilde v$. As a result, one obtains $v_c=v_{\mathcal R}\exp(-3D_v/2)$. In this study, the exponential factor yields a small correction 4\%-11\% to the cluster size.

The quantities $C$, $v_c$, and $\langle\mu\rangle$ can be resolved from the system of Eqs. (\ref{sysone}), (\ref{systwo}) and (\ref{systhree}). Further on, the mean cluster size will be referred to as a diameter relating to the mean cluster volume $v_c=\pi D_c^3/6$.

Thus, in this study, the neutron depolarization analysis adds to more complete quantitative characterization of the SPM system.

A constant value of the perovskite-cube volume $v_\circ=0.0558$ nm$^3$ was used for evaluation of the parameters in the relevant interval 97 K $<T<140$ K neglecting its relative variation of the order $1\cdot 10^{-3}$ due to thermal expansion [\onlinecite{r23}].
   
\section{Results and discussion}

The real [Figs. 1(a)-1(d)] and imaginary [Figs. 1(e)-1(h)] parts of the $M_2$ signal are presented as functions of the scanned magnetic field for four characteristic temperatures.
\noindent
\begin{figure}
\includegraphics[width=17cm]{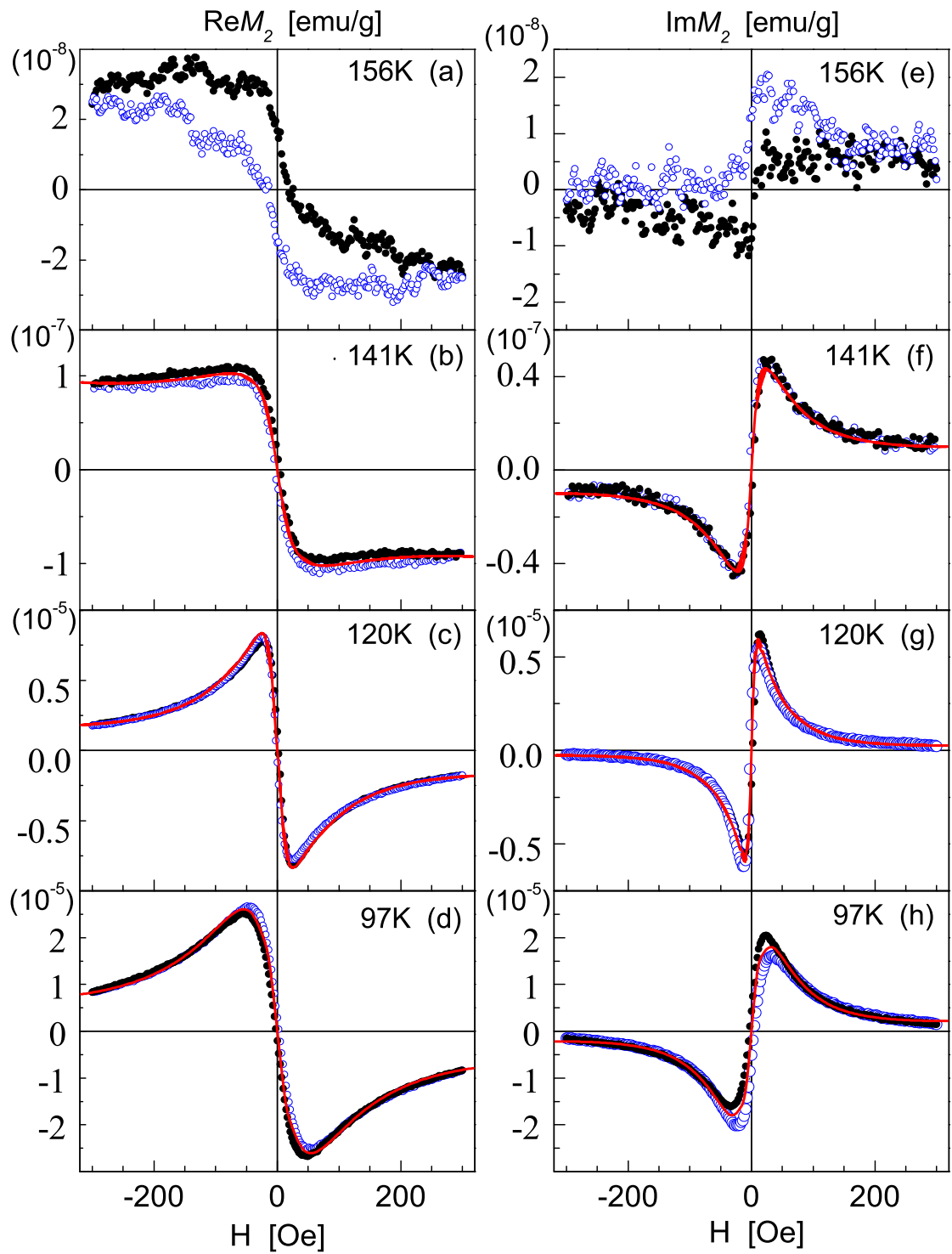}
\caption{\label{fig1}(Color online) The second harmonic of ac response as a function of dc magnetic field for four characteristic temperatures: the real part (a)-(d) and the imaginary part (e)-(h). Black solid and blue open circles denote direct and reverse scans, respectively. The red curves on the plots (b), (c), (d), (f), (g), and (h) are best fits (see text).}
\end{figure}

In Figs. 1(a) and 1(e), the signal for $T=156$ K is presented as a typical example of the response in the wide range 147 K $<T<230$ K. The regular part of the signal, noticeably contaminated with the apparatus noise, is suggested to originate from small-size magnetic inhomogeneities, which are, most likely, spin polarons and a small portion, $C\alt 10^{-4}$, of fine FM clusters located at spatial chemical inhomogeneities formed by oxygen nonstoichiometry and local enrichment with Sr, favorable for the clusters nucleation. A great deal of the response still extends far beyond the measured $H$-field range. The marked hysteresis is due to pinning of the clusters magnetic moments. At elevated temperatures, the signal gradually loosens and eventually fades in the apparatus noise somewhere near $T^*\approx 230$ K [\onlinecite{r13}].

At lower temperatures $T<147$ K, the character of the $M_2$ response, exemplified in Figs. 1(b) and 1(f) for $T=141$ K, drastically changes. In a temperature interval of only 15 K, both the real and imaginary parts grow rapidly by an order of magnitude. The hysteretic behavior becomes much less pronounced. Upon cooling down to 140 K, the \textquotedblleft coercive field\textquotedblright $H_{c2}$ defined by the condition Re$M_2(H_{c2})=0$ [Fig. 2, inset (a)] strongly falls. Such a form of the signal is typical for isolated SPM clusters. The extrema positions shift noticeably towards lower fields indicating the clusters growth, as evidenced by the calculations below.
\noindent
\begin{figure}
\includegraphics[width=17cm]{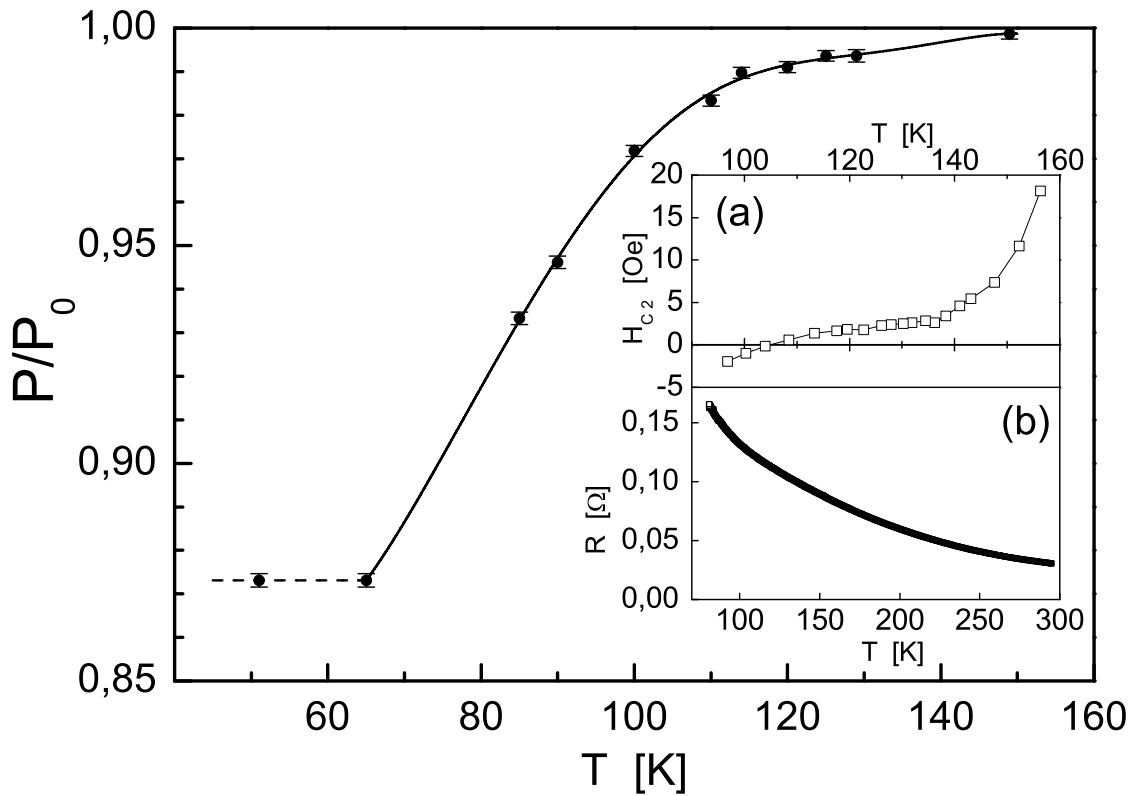}
\caption{\label{fig2}Normalized polarization of the neutron beam passed through the sample vs temperature. The solid curve is a polynomial fit. (Insets) \textquotedblleft Coercive field\textquotedblright estimated from the real parts of $M_2$ (a) and resistance of the sample (b) vs temperature.}
\end{figure}

Starting from 130 K, the rate of the signal growth from SPM clusters strongly increases. At this stage, illustrated by Figs. 1(c) and 1(g) for $T=120$ K, the signal becomes redistributed in favor of low fields exhibiting extrema at $H<50$ Oe in the real part and $H<30$ Oe in the imaginary part. The cluster signal dominates absolutely over the matrix response, the hysteresis being almost absent. This tendency persists down to the lowest temperatures measured.

Below 120 K [Fig. 1(d) and 1(h) for $T=97$ K], the signal continues to increase moderately with almost no change of its shape. This is the stage where the mean size and magnetic moment of clusters start to stabilize.

The parameters characterizing the cluster system at temperatures 97 K $<T<140$ K were evaluated from the best fits of $M_2(H)$ responses and the neutron depolarization data, as explained above. The latter are displayed on Fig. 2, where the polarization of the beam passed through the sample normalized by the polarization of the incident beam is presented as a function of temperature. The error bars are close to the size of points. The solid curve is a polynomial fit, to interpolate between the points. Upon lowering the temperature, depolarization of the passed beam increases, due to growth of the FM-cluster ensemble. At $T_f\approx 65$ K, the polarization ceases to fall and stabilizes at a constant level, indicating the end of FM-cluster evolution and freezing the FM-cluster dynamics on the time scale at least $10^4$ s. A value of the freezing temperature $T_f$ is compatible with NMR data [\onlinecite{r6}] and off-site ac MS measurements [\onlinecite{r36}]. A weak inflection between 130 and 150 K correlates with the appreciable growth of FM clusters below 140 K. The descending temperature dependency of resistivity [Fig. 2, inset (b)] evidences a dielectric character of the sample, typical for the doping $x<x_c$.

\subsection{Magnetic and geometrical characteristics}

In Fig. 3, some quantities obtained solely from nonlinear magnetic response are presented as functions of temperature. These are the saturation magnetization of the cluster ensemble $M$ and the characteristic dipolar energy $\varepsilon_d$ (a), the mean cluster magnetic moment $m_c$ and the cluster moment $m_m$ corresponding to the maximum position of the volume log-normal distribution (b), the concentration of clusters $N$ and the mean intercluster distance $\langle r\rangle$ (c), the anisotropy field $H_a$ and the mean cluster anisotropy energy $\varepsilon_a$ (d), and the width of the log-normal distribution function (e).
\noindent
\begin{figure}
\includegraphics[width=17cm]{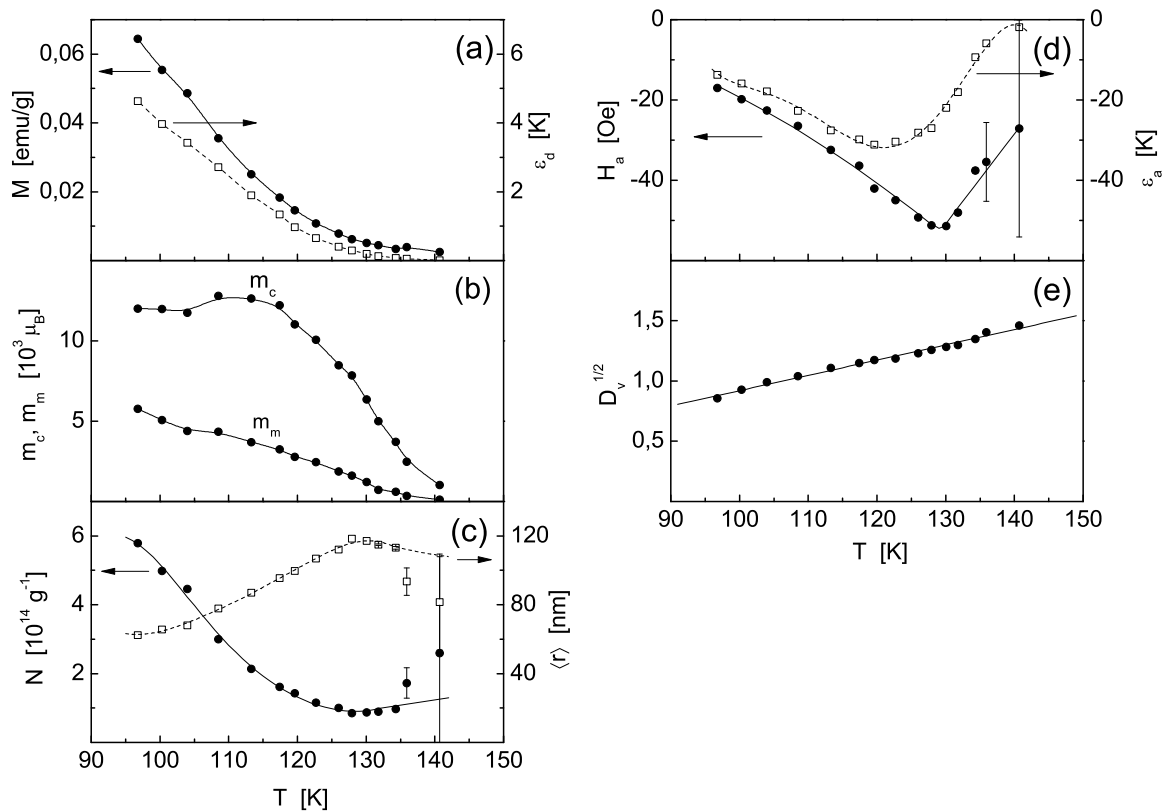}
\caption{\label{fig3}The total saturation magnetization of the clusters ensemble (solid circles) and the characteristic dipolar energy associated with the cluster subsystem (open squares) (a); the mean magnetic moment $m_c$ and the magnetic moment $m_m$ corresponding to a maximum position of the volume distribution (b); the concentration of clusters (solid circles) and the mean intercluster distance (open squares) (c); the incluster anisotropy field (solid circles) and the mean cluster anisotropy energy (open squares) (d), and the width of the log-normal distribution function (e) as functions of temperature. The solid curves in (a)-(d) are guides for the eye; the line in (e) is the linear approximation $\sqrt{D_v}\approx -0.349+0.0127T$.}
\end{figure}

Temperature dependencies of the parameters obtained by means also of the neutron polarization data are presented in Fig. 4, namely, the mean cluster diameter $D_c$ and the diameter $D_m$ corresponding to a maximum of the volume distribution (a), the volume fraction of clusters $C$ (b), and the mean magnetic moment per formula unit in the clusters $\langle\mu\rangle$ (c).

These dependencies specify the tendencies exemplified in Figs. 1(b)-1(d) and 1(f)-1(h) and visualize two stages of the SPM cluster MEPS alternating each other upon cooling, namely, (i) growth of FM clusters bound to chemical inhomogeneities in the temperature range down to 130 K and (ii) homogeneous nucleation of clusters below this temperature. The latter stage, in turn, proceeds via two regimes, viz., progressive increase of the mean cluster size down to 115 K and its stabilization below this temperature, the concentration of clusters still increasing.
\noindent
\begin{figure}
\includegraphics[width=17cm]{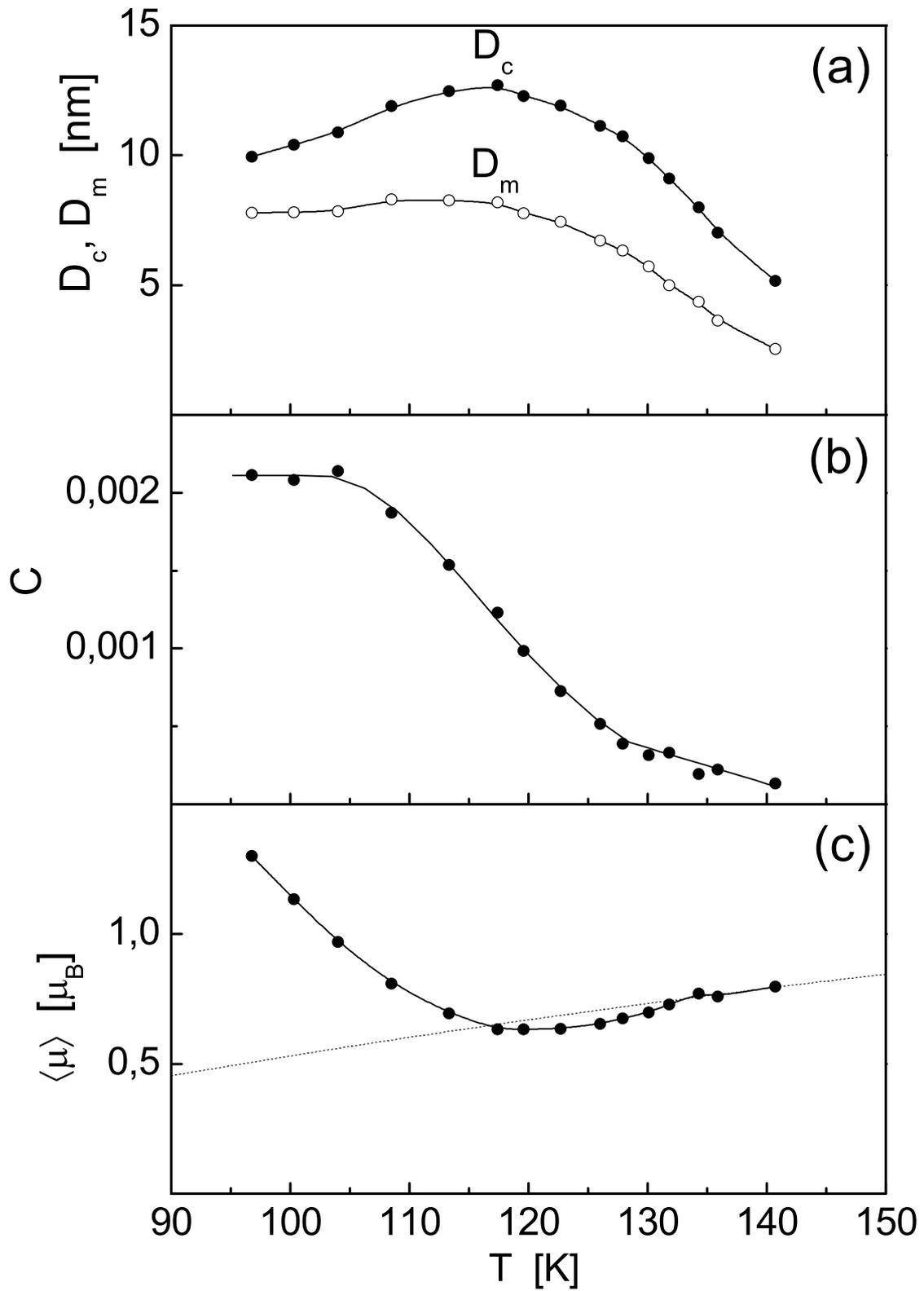}
\caption{\label{fig4}The mean cluster diameter $D_c$ and the diameter corresponding to a maximum of the volume distribution $D_m$ (a), the volume fraction of clusters (b), and the mean incluster magnetic moment per formula unit (c) as functions of temperature. The solid curves are guides for the eye; the dashed curve in (c) is $\mu_{eff}\exp(-\Delta/T)$ with $\mu_{eff}=2.15\mu_B$ and $\Delta=140$ K.}
\end{figure}

Now, these processes will be considered in more detail. At the first stage, growth of FM clusters bound to chemical inhomogeneities occurs gradually upon cooling [Figs. 3(b) and 4(a)]. At $T\approx 140$ K, the clusters become appreciably larger than the chemical inhomogeneities. The pinning of the cluster magnetic moment strongly weakens resulting in reduction of magnetic hysteresis [Figs. 1(b) and 1(f)] evidenced by substantial lowering the $M_2$ \textquotedblleft coercive field\textquotedblright down to $H_{c2}\approx 2$ Oe [Fig. 2, inset (a)]. Thus, $T_s=140$ K may be accepted as the onset of SPM behavior in the cobaltite under study. At this temperature, the cluster magnetic moment exceeds 10$^2\mu_B$ (150 formula units), large enough for the SPM formalism to become applicable. However, the clusters themselves still remain bound down to 130 K. The concentration of clusters and the mean intercluster distance [Fig. 3 (c)] at these temperatures vary only a little, if at all. The former may be considered as an estimate for the concentration of chemical inhomogeneities responsible for the initial cluster formation, $8.5\cdot 10^{14}$ cm$^{-3}$, with the mean distance $\langle r \rangle\approx 100$ nm between them. Slow increase of the saturation magnetization [Fig. 3(a)] and cluster fraction [Fig. 4(b)] above 130 K occurs only due to the growth of the cluster size.

At this stage, the mean magnetic moment per formula unit [Fig. 4 (c)] obeys an exponential law $\langle\mu\rangle=\mu_{eff}\exp(-\Delta/T)$ with $\Delta=140$ K and $\mu_{eff}=2.15\mu_B$ (dashed curve). Such kind of behavior, found in pure LaCoO$_3$ by electron spin resonance for not too high temperatures [\onlinecite{r20}], is conventionally interpreted as thermal excitation to the intermediate-spin state or, alternatively, a mixture of low-spin- and high-spin states, valid for not too high temperatures. Thus, above 130 K, $\langle\mu\rangle$ of the doped cobaltite under study also follows the common tendency of thermal excitation, predominantly, of Co$^{3+}$ ions, implying that the incluster hole concentration  only a little exceeds the concentration of holes in the matrix, $x\approx 0.15$. In this temperature region, the concentration of holes in the clusters is independent of temperature and determined, mainly, by the distribution of chemical inhomogeneities.

Below $T\approx 130$ K, the cluster fraction [Fig. 4(b)] and the saturation magnetization [Fig. 3(a)] start to rise steeply signifying intense growth of the sample volume involved into the SPM component. The concentration of clusters increases [Fig. 3(c)] manifesting entering the stage of free condensation (or homogeneous nucleation). The clusters are, mainly, no longer bound to chemical inhomogeneities and formation of new free clusters is activated over the whole sample volume resulting in increase of $N$ by five times upon cooling to 97 K. This transition is most distinctly visualized via a cusp in the temperature behavior of the anisotropy field [Fig. 3(d)]. The \textquotedblleft negative\textquotedblright sign of $H_a$ is consistent with the \textquotedblleft easy-plane\textquotedblright character of magnetocrystalline anisotropy found by neutron diffraction in this compound [\onlinecite{r23,r26}].

At the stage of free condensation, the mean magnetic moment $\langle\mu\rangle$ deviates from the exponential law [Fig. 4(c)]. This peculiar nonmonotonous behavior of the Co magnetic moment may be explained by variation of the concentration of holes in clusters with temperature. At first, the concentration even slightly lowers, whereas below 115 K, holes start to intensely fill the clusters importing additional uncompensated electron spin $s=1/2$, thus increasing $\langle\mu\rangle$. Due to the small volume fraction of clusters, $C\alt 10^{-3}$, this process does not cause noticeable hole depletion of the matrix.

Two different regimes can be distinguished at the stage of free condensation. Upon cooling down to $T\approx 115$ K, the magnetic moment and cluster diameter strongly increase [Figs. 3(b) and 4(a), respectively], while $\langle\mu\rangle$ even slightly decreases [Fig. 4(c)]. Below 115 K, $m_c$ and the cluster diameters $D_c$ and $D_m$ cease to grow reaching the large values $m_c\approx (12-13)\cdot 10^3\mu_B$ and $D_c\approx 12.5$ nm, whereas $\langle\mu\rangle$ rapidly rises. The position of the distribution maximum stabilizes at the value $D_m\approx 8$ nm (compare to the long-range FM correlation length $\ge 7$ nm evaluated by Phelan and collaborators [\onlinecite{r7}] in the polarized-neutron study of La$_{0.85}$Sr$_{0.15}$CoO$_3$ below 100 K). At the same time, the mean cluster diameter $D_c$ starts to decrease. The latter occurs due to gradual narrowing of the (asymmetrical) volume distribution of the cluster ensemble from both sides towards the maximum position. From $M_2$ data treatment, the distribution width varies linearly with temperature as $\sqrt{D_v}\approx -0.349+0.0127T$ [Fig. 3(e)]. Upon cooling below 115 K, the distribution \textquotedblleft sharpens\textquotedblright around the stabilized $D_m$. Thus, absorption of holes by clusters and increase of the clusters concentration are the main factors promoting the steady rise of the saturation magnetization [Fig. 3(a)] in the latter regime.

Charging the clusters due to absorption of holes increases the Coulomb energy of a cluster opposing its growth. The alternation of the regimes at 115 K is a result of the competition between these two tendencies.

Note also that Coulomb repulsion between the clusters may affect their space distribution especially at the lowest temperatures measured where the clusters are essentially enriched with holes.

However, particular reasons for the change of the MEPS character at 130 K as well as for alternation of the regimes at 115 K remain an open question.

By the way, poor convergence of the fit procedure at the high-temperature margin resulting from growing correlations between the parameters yields large error bars achieving 100\% for some parameters at 141 K [Figs. 3(c) and 3(d)]. The small cluster size implies a part of the system to be somewhat beyond the SPM formalism applicability limits and leads to serious convergence problems. Hence, quantitative results at temperatures close to 140 K should be referred to with care. 
\noindent
\begin{figure}
\includegraphics[width=17cm]{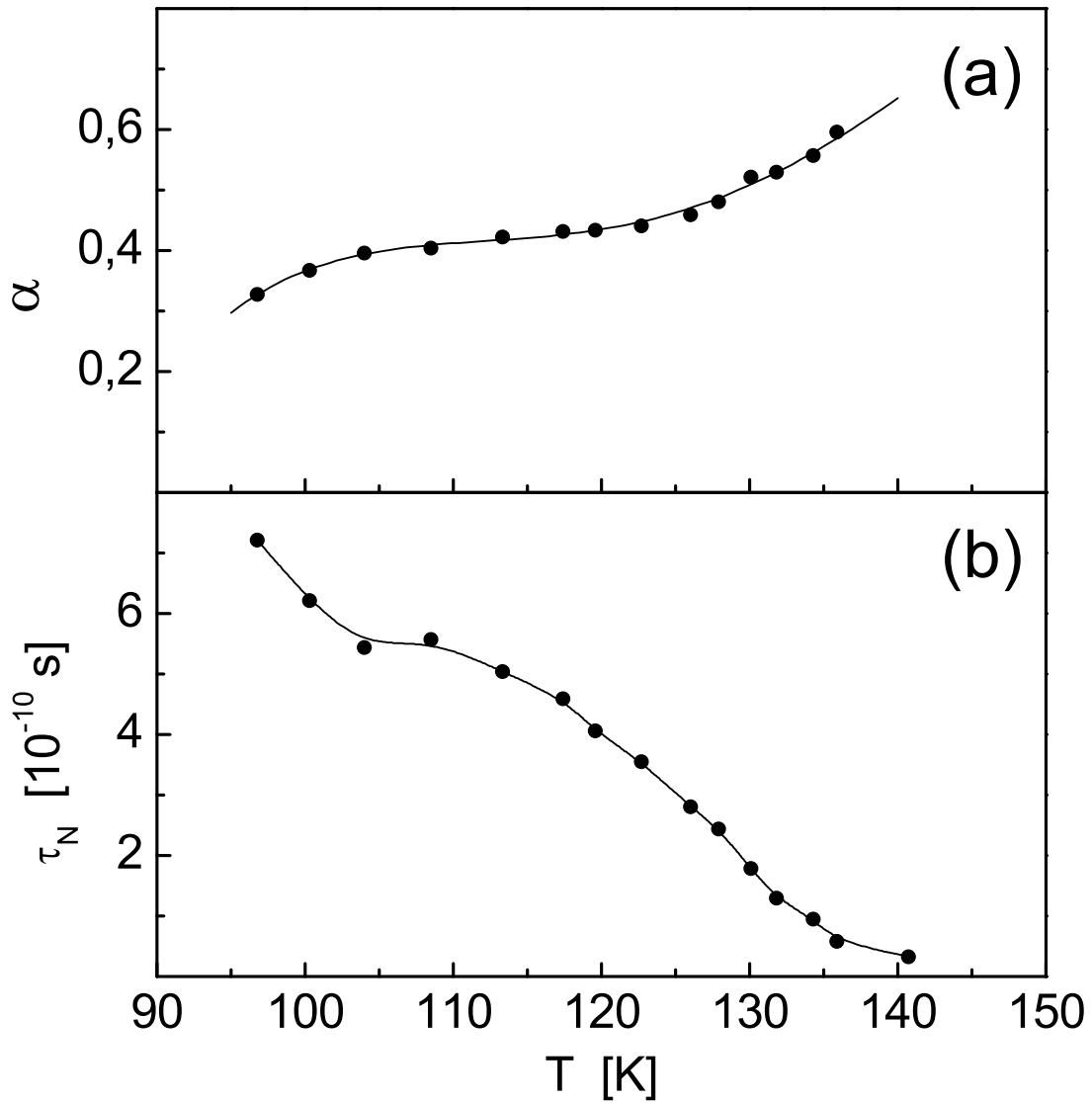}
\caption{\label{fig5} The damping factor (a) and the mean diffusion relaxation time of the cluster magnetic moment (b) as functions of temperature. The solid curves are guides for the eye.}
\end{figure}

\subsection{Superparamagnetic dynamics}

When studying an SPM system, a conventional problem arises on the effect of interparticle interactions on SPM dynamics [\onlinecite{r39}]. For a diluted SPM system with the insulating matrix (the present case), the most relevant is the long-range dipolar coupling between the particles. Its role can be estimated comparing the characteristic dipolar energy $\varepsilon_d$ [Fig. 3(a)] to the mean anisotropy energy of a cluster $\varepsilon_a$ [Fig. 3(d)]. The modulus of the latter well exceeds $\varepsilon_d$ except only for the lowest measured temperatures, thus, justifying disregard of dipolar forces in the calculations. At the margin $T=97$ K, however, $|\varepsilon_a|$ is only three times greater than $\varepsilon_d$. Below 97 K, the cluster anisotropy energy tends to decrease (by modulus), while the dipolar energy increases. The ratio $\varepsilon_d/|\varepsilon_a|$ grows upon cooling signifying the dipolar forces coming gradually into play. Provided that this trend persists at lower temperatures, the two quantities meet each other somewhere above the freezing temperature $T_f$. In this case, the FM-cluster subsystem enters the regime of dipolar dynamics favoring intercluster long-range magnetic correlations and inhibiting correlations with the anisotropy axes directions. The SPM dynamics acquires a collective character resulting, particularly, in increasing the characteristic relaxation time and the blocking temperature [\onlinecite{r40,r41,r42}]. Eventually, the cluster subsystem freezes in the dipolar-cluster-glass (or super spin-glass [\onlinecite{r41}]) state.

The damping factor $\alpha$ and the relaxation time $\tau_N$ (Fig. 5) are also deduced only from $M_2$ measurements. The former is a purely dynamical quantity, hardly measurable with other techniques. This parameter is responsible for relaxation of the FM-cluster magnetic moment due to its interaction with the environment. This might be incluster lattice and magnetic excitations, magnetic inhomogeneities, etc. To our knowledge, there is no microscopic theory of SPM relaxation. In the formalism, $\alpha$ was assumed to be explicitly dependent neither on magnetic field, nor on any cluster-system parameter. Its smooth rise with temperature is characteristic of a quantity depending on thermal excitations. The steeper increase with temperature above 130 K may be due to the growing effect of magnetic inhomogeneities as the clusters become bound and their magnetic moments pinned to chemical inhomogeneities as their size decreases. The steeper decrease of $\alpha$ below 115 K is suggestive of an onset of the freezing tendency enhanced (or provoked) by the intercluster dipolar forces. The effect of dipolar forces on the quantity characterizing free diffusion should not confuse, as the latter was defined with respect to the magnetic potential (\ref{two}) not containing intercluster interactions. Explicit inclusion of the dipolar-coupling term into Eq.~(\ref{two}) would have eliminated this effect from $\alpha$ and $\tau_N$ by transferring it to the potential-dependent part, such as an exponential in the N{\'e}el relaxation time in its conventional definition in the case $\sigma>0$.

A magnitude of the damping factor itself yields essential information concerning the influence of precession on relaxation of the cluster magnetic moment. The first and the second terms in the right-hand side of Eq.~(\ref{one}) describe precession and thermal, diffusion-type, relaxation of the magnetic moment, respectively. In the case of axial symmetry, when the magnetic field is parallel to the anisotropy axis, as well as in the limit $\alpha\gg 1$ (overdamped case), the first (precession) term vanishes in favor of diffusion. The measured values $\alpha=0.3-0.6$ point out that relaxation is considerably modified by precession. Both the terms of FPE turn to be comparable generating interplay of precession and thermal diffusion. Upon cooling, the role of precession increases. This phenomenon was examined in simulations by direct numerical solution of the Landau-Lifshitz-Gilbert equation (Ref.~[\onlinecite{r17a}] and references therein).

The relaxation time $\tau_N$ [Fig. 5(b)] experiences intense increase upon cooling following the combined effect of $m_c$, the damping factor, and the temperature, with the main contribution from the growing cluster magnetic moment. The values $\tau_N$ of the order $10^{-10}$ s are quite typical of SPM dynamics. The steeper growth below 115 K is suggestive of the beginning of the slowing-down tendency leading, eventually, to freezing the cluster system.
\\
\noindent
\begin{figure}
\includegraphics[width=17cm]{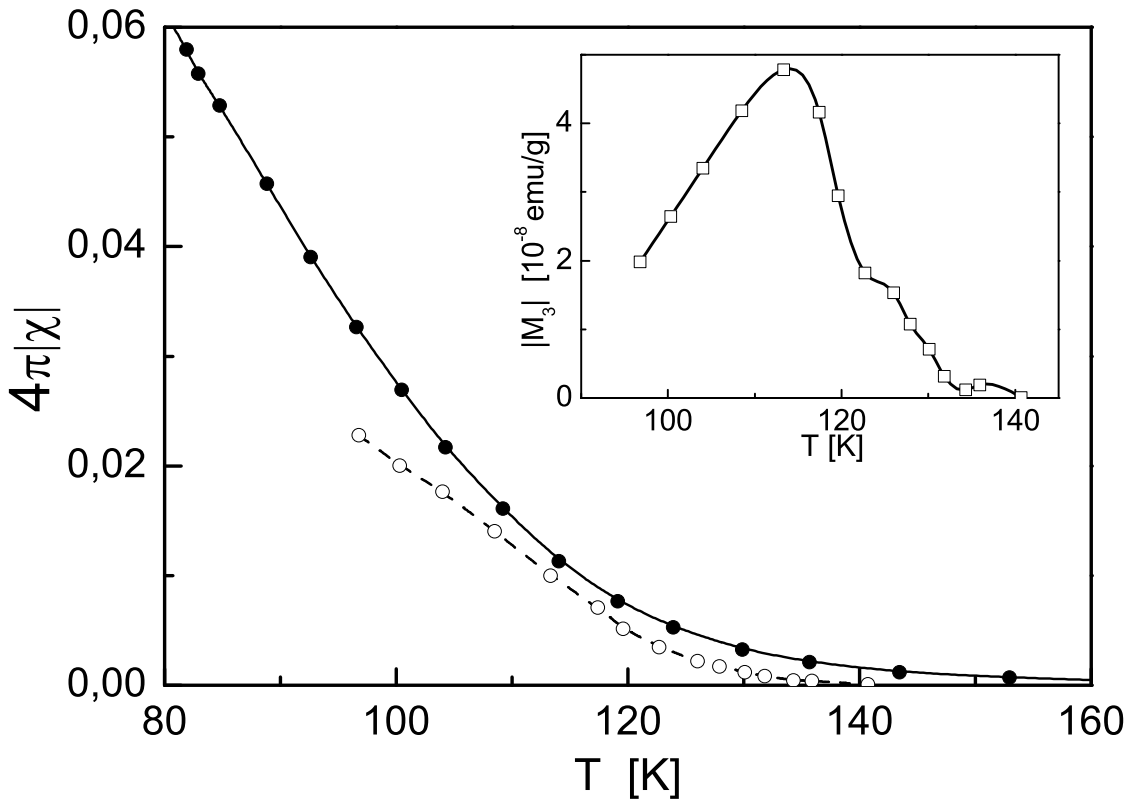}
\caption{\label{fig6}The measured total linear magnetic susceptibility (solid circles and line) and the clusters linear susceptibility as recovered from the second harmonic data for the excitation frequency 95 kHz (open circles and dashed line). (Inset) Third harmonic of the response vs temperature recovered for the excitation frequency 30 kHz. The curves are guides for the eye.}
\end{figure}

\subsection{Linear and third-order susceptibilities}

With known parameters of the SPM system, it is possible to retrieve the linear magnetic response of FM clusters and to estimate its role in magnetism of the cobaltite under study. In Fig. 6, temperature dependencies of the total magnetic susceptibility measured at 95 KHz and zero dc magnetic field and the linear SPM response recovered to these experimental conditions from $M_2$ measurements are presented. The recovered SPM susceptibility arises at $T\approx 140$ K correlating with the onset of FM clusters growth [Figs. 3(b) and 4(a)] and SPM behavior. Upon cooling, its relative contribution increases and becomes dominant below 120 K, in the region of intense growth of the saturation magnetization $M$ [Fig. 3(a)]. Recall, that for all temperatures measured, the clusters occupy not more than $2\cdot 10^{-3}$ of the sample volume [Fig. 4(b)], indicating that the large cluster magnetic moment is the main factor highlighting the SPM contribution to magnetic susceptibility.

In the inset to Fig. 6, the third harmonic is also displayed as retrieved for zero dc field and the ac field with the frequency 30 kHz and the amplitude 1 Oe. Its temperature dependency with the broad maximum at $T\approx 115$ K stands alongside with the behavior of the low-frequency third harmonics measured by Caciuffo and collaborators on the powder La$_x$Sr$_{1-x}$CoO$_3$ for a set of doping values in the \textquotedblleft metallic\textquotedblright region where FM clusters distinctly exhibited themselves in a pure form [\onlinecite{r23}]. These authors observed qualitatively similar dependencies with the maxima at $T\approx 175$ and 150 K for $x=0.25$ and 0.20, respectively, correlating with the maxima in respective linear susceptibilities. For $x=0.15$, the linear response exhibited a wide hump in the interval 150 K $<T<200$ K associated with FM clusters (unfortunately, the third harmonic was not presented for this doping). The latter is absent in Fig. 6 due to much lesser concentration of clusters in the single crystal.

\section{SPM dynamics ansatz: Gilbert \lowercase{vs} Landau-Lifshitz}

High accuracy and representativity of the experimental data enables to put a question concerning the main alternative underlying the formalism used. 

Historically, the Gilbert dynamical equation describing evolution of the classical magnetic moment $\bf m$ of a large FM cluster was introduced phenomenologically as a starting point (unpublished work, mentioned in Ref.~[\onlinecite{r52}]) [\onlinecite{r17a}]:
\begin{equation}
\frac{{\rm d}{\bf m}}{{\rm d}t}=\alpha{\bf m}\times\bigg[{\bf B}_{eff}-\frac{\alpha}{\gamma m}\frac{{\rm d}{\bf m}}{{\rm d}t}\bigg]~,
\label{seven}
\end{equation}
where the effective field ${\bf B}_{eff}$ is given by
\[
{\bf B}_{eff}=-\frac{\partial V}{\partial\bf m}
\]
with $V$ from Eq.~(\ref{two}). Equation ({\ref{seven}) contains time derivatives of the magnetic moment in both the right- and left-hand sides. This inconvenience was eliminated by expressing the derivative explicitly, resulting in the Landau-Lifshitz equation [\onlinecite{r17a,r53}]:
\begin{equation}
\frac{{\rm d}{\bf m}}{{\rm d}t}=\tilde\gamma{\bf m}\times{\bf B}_{eff}-\tilde\gamma\frac{\alpha}{m}{\bf m}\times[{\bf m}\times{\bf B}_{eff}]
\label{eight}
\end{equation}
with the renormalized gyromagnetic ratio $\tilde\gamma=\gamma/(1+\alpha^2)$. The famous Landau-Lifshitz relaxation (damping) term proportional to $-{\bf m}\times[{\bf m}\times{\bf B}_{eff}]$ drives $\bf m$ to the direction of ${\bf B}_{eff}$, while $\alpha$ measures the magnitude of the relaxation term relative to the gyromagnetic term.

In this transformation, no restrictions or additional assumptions were involved; thus, both the equations, (\ref{seven}) and (\ref{eight}), are of one and the same level of generality. Moreover, both of them lead, formally, to one and the same FPE; hence, each of the equations may be considered as generating one for the FPE. At first sight, the choice of the commencing equation seems to be a scholastic question. However, due to renormalization of $\gamma$, the characteristic diffusion time $\tau_N$ entering Eq.~(\ref{one}) is different in these two cases, depending on whether Gilbert- or Landau-Lifshitz equation is chosen as commencing one. The diffusion time is $\tau_N=\tau_\circ(\alpha+\alpha^{-1})$ in the former and $\tau_N=\tau_\circ\alpha^{-1}$ in the latter case. A question arises, in which scheme, Gilbert or Landau-Lifshitz, $\gamma$ has to be taken as bare one. Generally, this difference may result in dissimilar parameter meanings evaluated from best-fit calculations.

One might try to distinguish between the two schemes and to make a choice in favor of one of them by comparing chi-squares of the experimental-data best fits. As mentioned above, in the overdamped case (large $\alpha$), the precession term in Eq.~(\ref{one}) eliminates and the solution depends on the damping factor only implicitly via $\tau_N$. Thus, one can hardly expect any meaningful difference in the chi-squares. In the opposite case, $\alpha\ll 1$, there is also no difference, as $\tau_N$ is the same for both the schemes. The presently obtained values of $\alpha$ [Fig. 5(a)] lie just in the intermediate region for which the search of distinction in the best-fit quality makes sense.

\begingroup
\begin{table}
\caption{\label{table}Normalized chi-squares of the best fits, the damping factor $\alpha$, the relaxation time $\tau_N$, and the anisotropy field $H_a$ evaluated for a set of temperatures for the cases of Landau-Lifshitz- and Gilbert schemes.}
\begin{ruledtabular}
\begin{tabular}{c@{\hspace*{1cm}}cccc@{\hspace*{1cm}}cccc}
&\multicolumn{4}{c}{Landau-Lifshitz: $\tau_N=\tau_\circ\alpha^{-1}$}& \multicolumn{4}{c}{Gilbert: $\tau_N=\tau_\circ(\alpha+\alpha^{-1})$} \\
$T$ [K]&$\chi_{norm}^2$&$\alpha$&$\tau_N$ [10$^{10}$ s]&$H_a$ [Oe]&$\chi_{norm}^2$&$\alpha$&$\tau_N$ [10$^{10}$ s]&$H_a$ [Oe] \\[1 mm]\hline
97  & 2.88 & 0.33 & 7.2 & -17 & 2.33 & 0.36 & 6.5 & -21 \\
100 & 2.96 & 0.37 & 6.2 & -20 & 2.25 & 0.41 & 5.5 & -26 \\
104 & 2.07 & 0.40 & 5.4 & -23 & 1.48 & 0.44 & 4.9 & -33 \\
109 & 1.94 & 0.40 & 5.6 & -26 & 1.53 & 0.42 & 5.6 & -40
\end{tabular}
\end{ruledtabular}
\end{table}
\endgroup

Appropriate calculations were performed for a number of points in the lower-temperature region where the SPM system is well formed, with the steady fit convergence. The results, including the normalized chi-squares and the most sensitive parameters among those determined solely from the nonlinear measurements, viz., the damping factor, the relaxation time, and the anisotropy field, are presented in the table. It is seen that the Gilbert scheme yields systematically smaller values of $\chi^2_{norm}$ than these for the opposite, Landau-Lifshitz, one. Thus, the Gilbert approach seems to be more realistic, at least, for the object under study. The two parameters shown in the table differ essentially for the two cases, while the differences between other parameters (not presented) are of the order 1\%. At any rate, calculations performed within either of the alternative schemes would result in qualitatively similar temperature dependencies.

To establish the extent of universality of this finding, a study of the wide set of more conventional SPM systems is needed. 

\section{Conclusion}

The study of La$_{0.85}$Sr$_{0.15}$CoO$_3$ in the wide temperature range from 97 to 230 K was performed with linear and nonlinear ac magnetic response as well as neutron depolarization techniques. The system of large ferromagnetic clusters was found to emerge as a particular type of magnetoelectronic phase separation, besides well-established spin polarons. The experimental data were treated with a thorough formalism based on the solution of the Fokker-Planck equation to obtain magnetic, geometrical, and dynamical parameters characterizing the cluster ensemble.

Evolution of the cluster system upon cooling was found to occur via two processes, namely, (i) gradual growth of clusters bound to local chemical inhomogeneities in the wide temperature range down to 130 K with reaching the superparamagnetic regime at 140 K accompanied by failure of pinning the cluster magnetic moment and (ii) free condensation (or homogeneous nucleation) below 130 K consisting in intense formation and growth of clusters over the whole sample volume accompanied by variation of the hole density in the clusters. The latter process proceeds in two regimes alternating each other at 115 K, viz., growth of the cluster size with the incluster hole content only slightly exceeding the nominal doping $x=0.15$ above this temperature and stabilization of the cluster size accompanied by intense increase of the magnetic moment per unit cell due to enrichment of clusters with holes below 115 K.

At the onset of superparamagnetic behavior, ferromagnetic clusters start to contribute noticeably to the total linear magnetic susceptibility and, below 120 K, their contribution becomes dominant, at least down to 97 K.

The growing effect of intercluster dipolar coupling upon cooling may result in a dipolar character of the frozen state below 65 K.

The observed ferromagnetic clusters are suggested to be (i) a precursor of the ferromagnetic state at higher doping levels and (ii) an onset of large ferromagnetic clusters revealed by neutron scattering below 100 K [\onlinecite{r7}]. Thus, the phase separation and its evolution in a single crystal at temperatures higher than 100 K  was shown to have a more complicated character than was supposed before.

Relaxation dynamics of the cluster magnetic moment was found to be, essentially, the interplay of precession and thermal diffusion. Below 115 K, the magnetic dynamics exhibits a tendency to freezing favored by intercluster dipolar coupling.

High representativity and accuracy of the ac measurements performed made it possible to elucidate the ground alternative underlying superparamagnetic dynamics and to make a choice, though not robust enough, in favor of Gilbert approach.

The method employed was demonstrated to be an efficient means for studying the systems exhibiting superparamagnetic behavior.

\begin{acknowledgments}

One of the authors (V.\,V.\,D.) greatly appreciates kind assistance from A.~Kiryanov in acquiring skills in usage of the PNPI computer cluster. We are grateful also to Yu.~Chernenkov for determining orientations of the sample crystallographic axes and to P.~Molkanov for help with $M_2$ measurements.
\end{acknowledgments}

\appendix*
\section{}

The formalism used is presented in Refs.~[\onlinecite{r1,r2,r5}]. By expanding the distribution function in Eq.~(\ref{one}) in the series over spherical harmonics,
\begin{equation}
W(t,\vartheta,\varphi)=\sum_{l=0}^\infty\sum_{m=-l}^lc_{lm}(t)Y_{lm}(\vartheta,\varphi)~, \label{A1}
\end{equation}
the problem is reduced to the recurrence relation [\onlinecite{r1,r2}],
\begin{equation}
{\bf S}_n=-[{\bf Q}_n+{\bf Q}^+{\bf S}_{n+1}{\bf Q}_{n+1}]^{-1}~, \label{A2}
\end{equation}
relative to the matrix continued fraction ${\bf S}_n$. The tridiagonal supermatrices ${\bf Q}_n$, ${\bf Q}_n^\pm$ are given by
\begin{eqnarray*}
{}[{\bf Q}_n]_{l,m}=\delta_{l-1,m}{\bf p}_n+\delta_{l,m}{\bf q}_n(m\omega)+\delta_{l+1,m}{\bf p}_n~, \\
{}[{\bf Q}_n^\pm]_{l,m}=\delta_{l-1,m}{\bf p}_n^\pm+\delta_{l,m}{\bf q}_n^\pm+\delta_{l+1,m}{\bf p}_n^\pm~,
\end{eqnarray*}
where ${\bf q}_n(m\omega)=-im\tau_N\omega{\bf I}+{\bf q}_n$, with ${\bf I}$ being the identity matrix. The supermatrices ${\bf p}_n$, ${\bf p}_n^\pm$, ${\bf q}_n$, and ${\bf q}_n^\pm$ are presented as follows [\onlinecite{r1}]:
\begin{eqnarray*}
{\bf p}_n^-=\pmatrix{{\bf 0}&{\bf 0}\cr{\bf d}_{2n-1}&{\bf 0}}~,\quad & {\bf p}_n^+=\pmatrix{{\bf 0}&{\bf b}_{2n}\cr{\bf 0}&{\bf 0}}~, \\
{\bf p}_n=\pmatrix{{\bf a}_{2n}&{\bf d}_{2n}\cr{\bf b}_{2n-1}&{\bf a}_{2n-1}}~,\quad & {\bf q}_n^-=\pmatrix{{\bf V}_{2n}&{\bf 0}\cr{\bf W}_{2n-1}&{\bf V}_{2n-1}}~, \\
{\bf q}_n^+=\pmatrix{{\bf Z}_{2n}&{\bf Y}_{2n}\cr{\bf 0}&{\bf Z}_{2n-1}}~,\quad & {\bf q}_n=\pmatrix{{\bf X}_{2n}&{\bf W}_{2n}\cr{\bf Y}_{2n-1}&{\bf X}_{2n-1}}~.
\end{eqnarray*}
Dimensions of the supermatrices ${\bf p}_n\,({\bf q}_n)$, ${\bf p}_n^+\,({\bf q}_n^+)$, and ${\bf p}_n^-\,({\bf q}_n^-)$ are $8n\times 8n$, $8n\times 8(n+1)$ and $8n\times 8(n-1)$, respectively. Their matrix elements are, in turn, the matrices with the elements given by
\begin{widetext}
\begin{eqnarray}
({\bf a}_l)_{n,m}=\delta_{n-1,m}a_{l,-l+m}^-+\delta_{n,m}a_{l,-l+m-1}+\delta_{n+1,m}a_{l,-l+m-2}^+~,\nonumber \\
({\bf b}_l)_{n,m}=\delta_{n,m}b_{l,-l+m-1}^-+\delta_{n+1,m}b_{l,-l+m-2}+\delta_{n+2,m}b_{l,-l+m-3}^+~,\nonumber \\
({\bf d}_l)_{n,m}=\delta_{n-2,m}d_{l,-l+m+1}^-+\delta_{n-1,m}d_{l,-l+m}+\delta_{n,m}d_{l,-l+m-1}^+~,\nonumber \\
({\bf X}_l)_{n,m}=\delta_{n-1,m}x_{l,-l+m}^-+\delta_{n,m}x_{l,-l+m-1}+\delta_{n+1,m}x_{l,-l+m-2}^+~,\label{A5} \\ 
({\bf Y}_l)_{n,m}=\delta_{n,m}y_{l,-l+m-1}^-+\delta_{n+1,m}y_{l,-l+m-2}+\delta_{n+2,m}y_{l,-l+m-3}^+~,\nonumber \\
({\bf W}_l)_{n,m}=\delta_{n-2,m}w_{l,-l+m+1}^-+\delta_{n-1,m}w_{l,-l+m}+\delta_{n,m}w_{l,-l+m-1}^+~,\nonumber \\
({\bf Z}_l)_{n,m}=\delta_{n+2,m}z_{l,-l+m-3}~,\quad\nonumber
({\bf V}_l)_{n,m}=\delta_{n-2,m}v_{l,-l+m+1}~.\nonumber
\end{eqnarray}
\end{widetext}
The last five supermatrices are drawn explicitly in Ref.~[\onlinecite{r5}].

Using the direction cosines for the vectors ${\bf H}$ and ${\bf h}$,
\[
\gamma_1=\sin\psi\cos\phi~,\quad\gamma_2=\sin\psi\sin\phi~,\quad\gamma_3=\cos\psi~,
\]
and
\[
\gamma_1^\prime=\sin\psi^\prime\cos\phi^\prime~,\quad\gamma_2^\prime=\sin\psi^\prime\sin\phi^\prime~,\quad\gamma_3^\prime=\cos\psi^\prime~,
\]
respectively, the matrix elements in Eqs.~(\ref{A5}) can be written in the form [\onlinecite{r1,r5}],
\begin{eqnarray*}
a_{n,m}=-i\frac{m\xi_h\gamma_3^\prime}{4\alpha}~,\quad
a_{n,m}^+=-i\frac{\xi_h(\gamma_1^\prime-i\gamma_2^\prime)}{8\alpha}\sqrt{(n+m+1)(n-m)}~, \\
b_{n,m}=-\frac{\xi_h\gamma_3^\prime n}{4}\sqrt{\frac{(n+1)^2-m^2}{(2n+1)(2n+3)}}~,\quad
b_{n,m}^+=-\frac{\xi_h(\gamma_1^\prime-i\gamma_2^\prime)n}{4}\sqrt{\frac{(n+m+1)(n+m+2)}{(2n+1)(2n-1)}}~, \\
d_{n,m}=\frac{\xi_h\gamma_3^\prime(n+1)}{4}\sqrt{\frac{n^2-m^2}{(2n+1)(2n-1)}}~,\quad
d_{n,m}^+=-\frac{\xi_h(\gamma_1^\prime-i\gamma_2^\prime)(n+1)}{8}\sqrt{\frac{(n-m)(n-m-1)}{(2n+1)(2n-1)}}~, \\
x_{n,m}=\frac{\sigma[n(n+1)-3m^2]}{(2n-1)(2n+3)}-\frac{n(n+1)}{2}-i\frac{m\xi_H\gamma_3}{2\alpha}~,\quad
x_{n,m}^+=-i\frac{\xi_H(\gamma_1-i\gamma_2)}{4\alpha}\sqrt{(n+m+1)(n-m)}~, \\
y_{n,m}=-\biggl(\frac{\xi_H\gamma_3 n}{2}+i\frac{\sigma m}{\alpha}\biggr)\sqrt{\frac{(n+1)^2-m^2}{(2n+1)(2n+3)}}~,\quad
y_{n,m}^+=\frac{\xi_H (\gamma_1-i\gamma_2)n}{4}\sqrt{\frac{(n+m+1)(n+m+2)}{(2n+1)(2n+3)}}~, \\
w_{n,m}=\biggl(\xi_H\gamma_3\frac{n+1}{2}-i\frac{\sigma m}{\alpha}\biggr)\sqrt{\frac{n^2-m^2}{4n^2-1}}~,\quad
w_{n,m}^+=\frac{\xi_H (\gamma_1-i\gamma_2)(n+1)}{4}\sqrt{\frac{(n-m)(n-m-1)}{4n^2-1}}~, \\
z_{n,m}=-\frac{\sigma n}{2n+3}\sqrt{\frac{[(n+2)^2-m^2][(n+1)^2-m^2]}{(2n+1)(2n+5)}}~,\quad
v_{n,m}=\frac{\sigma(n+1)}{2n-1}\sqrt{\frac{(n^2-m^2)[(n-1)^2-m^2]}{(2n+1)(2n-3)}}~,
\end{eqnarray*}
with the additional relations,
\begin{eqnarray*}
a_{n,m}^-=-(a_{n,-m}^+)^*~,\quad
b_{n,m}^-=-(b_{n,-m}^+)^*~,\quad
d_{n,m}^-=-(d_{n,-m}^+)^*~,\\
x_{n,m}^-=-(x_{n,-m}^+)^*~,\quad
y_{n,m}^-=-(y_{n,-m}^+)^*~,\quad
w_{n,m}^-=-(w_{n,-m}^+)^*~.
\end{eqnarray*}

The stationary ac response can be calculated from the continued-fraction solution ${\bf S}_1$ of Eq.~(\ref{A2}), viz.,
\[
{\bf C}_1\equiv\left(\begin{array}{c}\vdots\cr
{\bf c}_1^{-2}(\omega)\cr{\bf c}_1^{-1}(\omega)\cr{\bf c}_1^0(\omega)\cr
{\bf c}_1^1(\omega)\cr{\bf c}_1^2(\omega)\cr\vdots
\end{array} \right)
=\frac{1}{\sqrt{4\pi}}{\bf S}_1\cdot\left(\begin{array}{c}\vdots\cr
{\bf 0}\cr{\bf p}_1^-\cr{\bf q}_1^-\cr{\bf p}_1^-\cr{\bf 0}\cr \vdots\end{array} \right)~, \\*
\]
where each element of ${\bf C}_1$ is a column vector,
\begin{equation}
{\bf c}_1^k(\omega)=\left(\begin{array}{l} c_{2,-2}^k(\omega)\cr
c_{2,-1}^k(\omega)\cr c_{2,0}^k(\omega)\cr c_{2,1}^k(\omega)\cr c_{2,2}^k(\omega)\cr c_{1,-1}^k(\omega)\cr c_{1,0}^k(\omega)\cr
c_{1,1}^k(\omega)\end{array}\right)\label{A7}~, \\*
\end{equation}
and
\begin{eqnarray*}
{\bf q}_1^-=\left(\begin{array}{c} 0\cr 0\cr\frac{\displaystyle 2\sigma}{\displaystyle\sqrt{5}}\cr 0\cr 0\cr\frac{\displaystyle(\gamma_1-i\gamma_2)\xi_H}{\displaystyle\sqrt{6}}\cr\frac{\displaystyle\gamma_3\xi_H}{\displaystyle\sqrt{3}}\cr
-\frac{\displaystyle(\gamma_1+i\gamma_2)\xi_H}{\displaystyle\sqrt{6}} \end{array}\right)~,\quad
{\bf p}_1^-=\left(\begin{array}{c} 0\cr 0\cr 0\cr 0\cr 0\cr
\frac{\displaystyle(\gamma_1^\prime-i\gamma_2^\prime)\xi_h}{\displaystyle 2\sqrt{6}}\cr \frac{\displaystyle\gamma_3^\prime\xi_h}{\displaystyle2\sqrt{3}}\cr
-\frac{\displaystyle(\gamma_1^\prime+i\gamma_2^\prime)\xi_h}{\displaystyle 2\sqrt{6}}\end{array}\right)~. \\*
\end{eqnarray*}
Elements of the column vector ${\bf c}_1^k(\omega)$ are Fourier components of the corresponding coefficients $c_{lm}$ in the expansion (\ref{A1}).

Following Refs.~[\onlinecite{r1,r2}], we define the ac response as magnetization $M_h(t)$ in the direction of the driving field {\bf h}, with the Fourier transform,
\[
M_h(t)=M_s\sum_{k=-\infty}^\infty m_1^k(\omega)e^{ik\omega t}~,
\]
where
\[
m_1^k(\omega)=\sqrt{\frac{4\pi}{3}}\biggl[\gamma_3^\prime c_{1,0}^k(\omega)+\frac{(\gamma_1^\prime +i\gamma_2^\prime)c_{1,-1}^k(\omega)- (\gamma_1^\prime -i\gamma_2^\prime)c_{1,1}^k(\omega)}{\sqrt{2}}\biggr]~,
\]
with $c_{1,0}^k(\omega)$, $c_{1,-1}^k(\omega)$, and $c_{1,1}^k(\omega)$ taken from Eq.~(\ref{A7}).

\bibliography{PaperLSCOnrevarch}

\end{document}